\documentclass[a4]{epl2}
\usepackage{graphicx}
\usepackage{epsfig}
\usepackage{color}
\usepackage{graphicx}
\title{\textcolor[rgb]{0.00,0.00,0.00}{Bohmian quantum mechanics revisited}}
\shorttitle{\textcolor[rgb]{0.00,0.59,0.00}{Bohmian quantum mechanics revisited}}

\author{\inst{}A. I. Arbab\inst{}\footnote{arbab.ibrahim@gmail.com}}

\institute{\inst{}
Department of Physics, College of Science, Qassim University, P.O. Box 6644, 51452 Buraidah,  KSA\\
\inst{}
Department of Physics,
Faculty of Science, University of Khartoum,  Khartoum 11115, Sudan}


\abstract{By expressing the Schr\"{o}dinger wave function in the form $\psi=Re^{iS/\hbar}$, where $R$ and $S$ are real functions, we have shown that the expectation value of $S$ is conserved. The amplitude of the wave ($R$) is found to satisfy the Schr\"{o}dinger equation while the phase ($S$) is related to the energy conservation. Besides the quantum potential that depends on $R$, \emph{viz.}, $V_Q=-\frac{\hbar^2}{2m}\frac{\nabla^2R}{R}$\,, we have obtained a phase  potential $V_S=-\frac{S\nabla^2S}{m}$ that depends on the phase $S$ derivative. The phase force is a dissipative  force.  The quantum potential may be attributed to the interaction between the two subfields $S$ and $R$ comprising the quantum particle. This results in splitting (creation/annihilation) of these subfields, each having a mass $mc^2$ with an internal frequency of $2mc^2/\hbar$, satisfying the original wave equation and endowing the particle its quantum nature. The mass of one subfield reflects the interaction with the other subfield. If in Bohmian ansatz  $R$ satisfies the Klein-Gordon equation, then $S$ must satisfies the wave equation. Conversely, if $R$ satisfies the wave equation, then $S$ yields the Einstein relativistic energy momentum equation.}

\begin{document}
\maketitle
\baselineskip=20pt
\hoffset=-1cm
\textwidth=18cm

\section{\textcolor[rgb]{0.00,0.07,1.00}{Introduction}}

To give a conceivable meaning for Schr\"{o}dinger quantum mechanics, Bohm had expressed the wavefunction representing the particle in the form $\psi=Re^{iS/\hbar}$\,, where $R$ and $S$ are two real functions \textcolor[rgb]{0.00,0.07,1.00}{\cite{bohm, bohm1,broglie}}. In doing so, Bohm had reformulated quantum mechanics by describing the particle classical trajectory by a quantum Hamilton-Jacobi equation while retaining the quantum aspect of the particle. Thus, his formulation includes some hidden variables that also describe the state of the particle. The particle is described by the action $S$  that satisfies the Hamilton-Jacobi equation, and $R$ satisfying the continuity equation. It is found that a quantum potential emerges from the Hamilton-Jacobi equation. This potential is assumed to guide the particle as it travels.

In this paper, we follow the same routine as Bohm but define another action-like quantity, that is $\hbar R$, associated with a different Hamiltonian.  Inasmuch as the ordinary Hamiltonian encompasses a quantum potential, the new Hamiltonian gives rise to a  classical  potential that we refer to phase  potential. The classical phase  potential is derived whose force yields a viscous force. The coefficient of viscosity is found to be directly proportional to the particle phase. In this formulation, the wave amplitude ($R$) is found to satisfy the Schr\"{o}dinger equation. The solution of this equation yields the spatial dependence of $R$, but the time dependence is obtained by solving the Hamilton equations of motion by treating $S$ and $R$ and their spatial derivatives ($\vec{\nabla}S$ and $\hbar\vec{\nabla}R$) as the corresponding generalized momenta. Solving the  Hamilton's equations for each system yields the relation between $R$ and $S$. In the Bohm formulation, the functions $R$ and $S$ are arbitrarily related functions. For a particular choice of the generalized co-ordinates ($S=2\pi\hbar\, R$), the quantum potential arising from $R$ curvature is equal to that arising from  $S$ curvature.

We apply the Bohm recipe to the Dirac equation, and the quantum telegraph equation that was recently shown to yield quantum mechanical identifications of quantum particles \textcolor[rgb]{0.00,0.07,1.00}{\cite{uqm}}. The amplitude ($R$) satisfies the original quantum telegraph equation, but the phase ($S$) satisfies the wave equation traveling at the speed of light. In a particular case, $S$ is shown to satisfy  the relativistic Einstein energy-momentum equation if $R$ satisfies a special telegraph equation.
A similar situation applies for the Klein-Gordon equation too. We treat here the functions $R$ and $S$, in the Klein-Gordon and the unified wave equations, as fields in which the wavefunction (field) is decomposed. They can also be thought of as eigen functions (basis) for the original wavefunction.

In this paper we present in Section 2 the standard Bohmian mechanics where we derived a new potential besides the quantum potential. In Section 3 we employ the canonical quantization by treating the probability conservation equations as defining another Hamilton-Jacobi equation. We then obtained the equation of motion of the two subfields $R$ and $S$ by treating them as generalized co-ordinates. We explore the Bohmian mechanics in Dirac formalism in Section 5. Section 6 and 7 are dealt with the Bohmian mechanics for the Klein-Gordon and the unified wave equations (the telegraph quantum equation). We end our paper by some concluding remarks.

\section{\textcolor[rgb]{0.00,0.07,1.00}{Bohmian quantum mechanics}}

Substituting $\psi=R\,e^{iS/\hbar}$ in the Schr\"{o}dinger's equation and equating the real and imaginary parts in the two sides of the resulting equation to each other, yield \textcolor[rgb]{0.00,0.07,1.00}{\cite{bohm}}
\begin{equation}
\frac{\partial S}{\partial t}=-\left(\frac{(\nabla S)^2}{2m}+V-\frac{\hbar^2}{2m}\frac{\nabla^2R}{R} \right) \,,
\end{equation}
and
\begin{equation}
\frac{\partial R}{\partial t}=-\frac{1}{2m}\left(R\nabla^2S+2\vec{\nabla}R\cdot\vec{\nabla}S\right) \,,
\end{equation}
the term $V_Q=-\frac{\hbar^2}{2m}\frac{\nabla^2R}{R}$ is known as the quantum potential. It represents the quantum effects observed in a system. It also reflects a curvature effect resulting from $R$. The presence of this term in the quantum Hamilton-Jacobi equation helps to view the quantum effects along a classical trajectory. The quantum potential also  acts to guide the movement of the quantum particle. The relation between $R$ and $S$ is obtained by expressing $\psi$ as $\psi=U+iW=R\cos(S/\hbar)+iR\sin(S/\hbar)$. This implies that $R^2=U^2+W^2$ and $S=\hbar\,\tan^{-1}(W/U)$. Notice that $U$ and $W$ satisfy the Schrodinger's equation, while $R$ and $S$ do not.

Let us now  multiply eq.(1) by $R^2$ and eq.(2) by $2RS$ and add the two resulting equations to obtain
\begin{equation}
\frac{\partial }{\partial t}\,(R^2S)+\frac{R^2(\nabla S)^2}{2m}+R^2V-\frac{\hbar^2}{2m}R\nabla^2R+\frac{R^2S\nabla^2S}{m}+\frac{2RS(\vec{\nabla}R\cdot\vec{\nabla}S)}{m}=0 \,.
\end{equation}
We further know that
\begin{equation}
\frac{d}{dt}(R^2S)=\frac{\partial (R^2S)}{\partial t}+\vec{v}\cdot\vec{\nabla}(R^2S)\,.
\end{equation}
But since $\vec{v}=\frac{\vec{\nabla}S}{m}$\,, then eq.(4) becomes
\begin{equation}
\frac{d}{dt}(R^2S)=\frac{\partial }{\partial t}(R^2S)+\frac{\vec{\nabla}S}{m}\cdot\vec{\nabla}(R^2S) \,.
\end{equation}
The phase of the classical wave is in general not a measurable quantity, but it can now have a physical meaning (measurable quantity). If this phase density is conserved, then the right hand-side of  eq.(5) vanishes, and hence one can write
\begin{equation}
\frac{\partial }{\partial t}\,(R^2S)+\frac{2RS(\vec{\nabla}S\cdot\vec{\nabla}R)}{m}+\frac{R^2(\nabla S)^2}{m}=0\,,
\end{equation}
by expanding the second term in the right hand-side of eq.(5).

Let us now combine eqs.(3) and (6) to obtain
\begin{equation}
\frac{\hbar^2}{2m}\frac{\nabla^2R}{R}-V=\frac{S\nabla^2S}{m}-\frac{(\nabla S)^2}{2m}\,.
\end{equation}
Now if we assume that $V$ is a function of $R$ only, one can solve eq.(7) by setting each side equal to a constant, $E$, \emph{viz.},
\begin{equation}
\frac{\hbar^2}{2m}\frac{\nabla^2R}{R}-V(R)=E\,,
\end{equation}
and
\begin{equation}
\frac{S\nabla^2S}{m}-\frac{(\nabla S)^2}{2m}=E\,.
\end{equation}
It is thus interesting that when $V$ is a function of $R$ only, then Eq.(7) decouples the quantum part from the classical one.
Equation (8) can be expressed as
\begin{equation}
-\frac{\hbar^2}{2m}\nabla^2R+VR=-ER\,,
\end{equation}
which is the Schr\"{o}dinger equation for  a particle of mass $m$ and total energy $-E$ whose wave function is $R$. Thus, unlike the Bohmian formulation, where $R$ doesn't satisfy the Schrodinger's equation, the function $R$ in our present formulation satisfies the Schrodinger's equation. Equation (9) is a non linear differential equation in the variable $S$ which can be solved to give the spatial dependance of $S$. Note that while Eq.(8) is a quantum equation, Eq.(9) is a classical one. Hence, our treatment of the Bohm's mechanics splits into classical as well as quantum equations. The amplitude is governed by a quantum equation, whereas the phase is governed by a classical one.

 The solution of eq.(9) in one dimension is
$$ \hspace{4cm} S=\frac{2mE}{C}+\frac{C}{4}\,x^2\,,\hspace{4cm}(a)$$
where $C$ is a constant having a dimension of $kg/s$.

Using the definition $\vec{p}=\vec{\nabla}S$, eq.(9) can be expressed as
\begin{equation}
\frac{p^2}{2m}-\frac{S\nabla^2S}{m}=-E\,.
\end{equation}
One can define the term
\begin{equation}
V_S=-\frac{S\nabla^2S}{m}\,.
\end{equation}
as a classical \emph{phase potential}. This can be juxtaposed with the quantum potential, $V_Q$,  that  depends on $\hbar^2$. This potential arises from an internal motion of the particle. Like the quantum potential, that is proportional to the curvature due to $R$, the phase  potential is proportional to the curvature  due to $S$.

Applying eq.(a) in eq.(12) yields $V_S=-E-(C^2/8m)\,x^2$.  This gives rise to a conservative force, $F_S=-\partial V_S/\partial x=(C^2/4m)\, x$. Such a potential is found to govern the nucleons (proton and neutron) inside a nucleus.

While the quantum potential is scale invariant ($R\rightarrow \beta R$, for some constant $\beta$), the phase potential is not and scales quadratically. That is because the amplitude $R$ can always be normalized to unity, but  $S$ is not. Moreover, while the quantum potential can be singular, the phase  potential is always regular and definite.

The force associated with the phase spatial variation in the potential is given by
$$
F_S=-\vec{\nabla}V_S=S\frac{\nabla^2(\vec{\nabla}S)}{m}+\frac{(\vec{\nabla}S)}{m}\nabla^2S\,,$$
or
\begin{equation}
F_S=S\,\nabla^2\vec{v}+(\nabla^2S)\,\vec{v}\,,
\end{equation}
where $\vec{v}=\frac{\vec{\nabla}S}{m}$. Since the second term in the force $F_S$ depends on the particle velocity, it can be seen as a drag force. The first term accounts for the viscosity force that is similar to the term  appearing in the Navier-Stokes equation. It may usher in a direction of an existence of a permeating fluid filling the whole space that were once proposed for Maxwell's waves (the ether). The second term can be expressed as $\nabla^2S=\vec{\nabla}\cdot(\vec{\nabla}S)$ which can be written as $m\vec{\nabla}\cdot\vec{v}$\,. Hence, the phase  force in eq.(13)  will read
\begin{equation}
F_S=S\,\nabla^2\vec{v}+m\vec{v}\,(\vec{\nabla}\cdot\vec{v})\,.
\end{equation}
Thus, because of its phase, the quantum particle would appear  to be  moving in a viscous fluid. The first term in eq.(14) suggests that the phase of the wavefunction of the particle  measures  some intrinsic  coefficient of viscosity, since the viscous force density is proportional to $\eta\nabla^2\vec{v}$, where $\eta$ is the coefficient of viscosity. Hence, $\eta$ represents the phase  density. Note that viscosity requires the presence of a fluid a priori. Equation (14) shows that the drag force depends on the particle mass (inertia), but the viscosity force does not. This amounts to say that it is associated with some internal degree of freedom that the particle has. And since, $S\ne0$, the viscous force can never vanish, unless $\nabla^2\vec{v}=0$. It is shown  by Schrodinger that the free electron, in Dirac's theory, exhibits a jittery motion. This motion may be connected with this force.

It is found recently that the Hall's viscosity is proportional to the average spin \textcolor[rgb]{0.00,0.07,1.00}{\cite{read}}.  Accordingly, one may associate the phase with the spin of the particle. In the realm of particle physics, hadrons consist of three quarks that move freely inside them. Consequently, one may argue that a similar situation exits inside the electron.

For an incompressible fluid, $\vec{\nabla}\cdot\vec{v}=0$, and therefore, $F_S=S\nabla^2\vec{v}$\,, which implies  that the phase  force is independent of the particle mass.

Let us digress a bit and look at eq.(11) as defining the energy conservation of a free particle ($V(R)=0$) being influenced by its internal motion only, \emph{viz.},
\begin{equation}
-E=\frac{p^2}{2m}+V_S\,,
\end{equation}
upon using eq.(12).
Using the vector identity
\begin{equation}
\vec{\nabla}\cdot(S\vec{\nabla}S)=(\nabla S)^2+S\nabla^2S\,,
\end{equation}
eq.(9) can be expressed as
\begin{equation}
\frac{3p^2}{2m}+E=\vec{\nabla}\cdot(S\frac{\vec{\nabla}S}{m})\,,
\end{equation}
or
\begin{equation}
\frac{3p^2}{2m}+E=\vec{\nabla}\cdot(S\vec{v})\,.
\end{equation}
However, if $S$ is conserved, then
\begin{equation}
\frac{dS}{dt}=\frac{\partial S}{\partial t}+\vec{\nabla}\cdot(S\vec{v})=0\,,
\end{equation}
so that eq.(18) can be written as
\begin{equation}
\frac{\partial S}{\partial t}+\left(\frac{3p^2}{2m}+E\right)=0\,.
\end{equation}
The factor 3 in the above equation may be attributed to rotation of the particle in 3-dimensions.
Equation (20) can be expressed in the Hamilton-Jacobi format to read
\begin{equation}
\frac{\partial S}{\partial t}+H_p=0\,,
\end{equation}
where
\begin{equation}
H_p=\frac{p^2}{2m}+\frac{p^2}{m}+E\,,
\end{equation}
is the Hamiltonian of the system. The extra kinetic energy term, $\frac{p^2}{m}$, may be attributed to spin kinetic energy contributed by the two subfields ($R$ and $S$) comprising the quantum particle.

\section{\textcolor[rgb]{0.00,0.07,1.00}{Canonical momenta}}

Let us express the momentum related to $S$ as $\vec{p}_S=\vec{\nabla}S$ and that is related to $R$ as $\vec{p}_R=\hbar\vec{\nabla}R$. Equations (1) and (2) can be expressed as
\begin{equation}
\frac{\partial S}{\partial t}+H_S=0\,, \qquad\qquad \frac{\partial {\cal R}}{\partial t}+H_R=0\,,
\end{equation}
where
\begin{equation}
H_S=\frac{(\nabla S)^2}{2m}+V-\frac{\hbar^2}{2m}\frac{\nabla^2R}{R} \,,
\end{equation}
and
\begin{equation}
H_R=\frac{\hbar}{2m}\left(R\nabla^2S+2\vec{\nabla}R\cdot\vec{\nabla}S)\right) \,.
\end{equation}
are the Hamiltonians associating with the two actions, $S$ and ${\cal R}=\hbar R$, respectively.

Thus, one can use $R$ and $S$ as generalized coordinates and $p_R$ and $p_S$ as their corresponding generalized momenta.  To make $R$ have a dimension of length and $S$ have a dimension of momentum, we multiple $R$ by a constant $\lambda$ and divide $S$ by a constant $p_\lambda$ having  dimensions of length and momentum, respectively, \emph{viz}.,  $\tilde{S}=\frac{S}{p_\lambda}$ and $\tilde{R}=\lambda R$.

In terms of the momenta $p_R$ and $p_S$\,,  eqs.(24) and (25) can be written as
\begin{equation}
H_S=\frac{p_S^2}{2m}+V-\frac{\hbar\lambda}{2m\tilde{R}}\,\vec{\nabla}.\vec{p_R}  \,,
\end{equation}
and
\begin{equation}
H_R=\frac{\hbar\,\tilde{R}}{2m\lambda\tilde{S}}\vec{\nabla}\cdot(\tilde{S}\vec{p}_S)-\frac{\hbar \tilde{R}\,p^2_S}{2m\lambda p_\lambda\tilde{S}}+\frac{\vec{p}_R\cdot\vec{p}_S}{m} \,.
\end{equation}
In this case, the Hamilton's equations become
\begin{equation}
\dot {\tilde{R}}=\frac{\partial H_R}{\partial p_R}\,,\qquad\qquad \dot p_R=-\frac{\partial H_R}{\partial \tilde{R}}\,,
\end{equation}
and
\begin{equation}
\dot{\tilde{S}}=\frac{\partial H_S}{\partial p_S}\,,\qquad\qquad \dot p_S=-\frac{\partial H_S}{\partial \tilde{S}}\,,
\end{equation}
 to solve for $\tilde{R}$ and $\tilde{S}$.
Apply eq.(28) in eq.(27) to obtain
\begin{equation}
\dot R=\frac{p_S}{\lambda m}\,,\qquad\qquad\qquad \dot p_R=\frac{\hbar\,p_S^2}{2mS\lambda }-\frac{\hbar}{2mS\lambda }\vec{\nabla}\cdot(S\vec{p}_S)\,,
\end{equation}
and apply eq.(29) in eq.(26) to obtain
\begin{equation}
\dot S=\frac{p_\lambda p_S}{m}\,,\qquad\qquad\qquad\qquad \dot p_S=-p_\lambda\frac{\partial V}{\partial S}\,.
\end{equation}
An educated guess for the relation between $\lambda$ and $p_\lambda$ is that $\lambda=h/p_\lambda$, where $h$ is the Planck's constant. If we assume that $V$ is a function of $R$ only, then $\dot p_S=0$. Hence, $ R$ and $ S$ are constants too. This yields the relation  $S=hR$. This is a very interesting relation relating the  wave amplitude to its phase. In this particular case, $V_S=V_Q$.

Recall that writing the wavefunction in the form $\psi=R\,e^{iS/\hbar}$ may be seen as defining $\psi$ as a point in a complex plane, \emph{viz.}, $z=re^{i\theta}$, where $r^2=|z|^2$ and $\theta$ is the angle that $r$ subtends with the x-axis. In this case, the relation $S=hR$ implies that $S/\hbar=2\pi R$ is the circumference of the circle with radius $R$. Winding the circle $n$ times, for some integer $n$, doesn't change the phase (we come to the same point we started with). Hence, we can say that $S=n\hbar$ for constant $R$.

\section{\textcolor[rgb]{0.00,0.07,1.00}{Phase-dependent potential}}

Let us now assume that $V$ depends on $S$. In this case eq.(7) reads
\begin{equation}
\frac{\hbar^2}{2m}\frac{\nabla^2R}{R}=-E_S\,,
\end{equation}
and
\begin{equation}
-E_S=\frac{S\nabla^2S}{m}-\frac{(\nabla S)^2}{2m}+V(S)\,.
\end{equation}
Equations (32) and (33) can be written as
\begin{equation}
\nabla^2R+\frac{2mE_S}{\hbar^2}R=0\,,
\end{equation}
and
\begin{equation}
E_S=\frac{p_S^2}{2m}+V_T\,,\qquad\qquad V_T=-\left(V+\frac{S\nabla^2S}{m}\right)\,,
\end{equation}
where $V_T$ is the total potential of the system. Therefore, the amplitude $R$ satisfies the Schr\"{o}dinger equation for a particle with total energy equals to $E_S$.

\section{\textcolor[rgb]{0.00,0.07,1.00}{Bohmian Dirac's equation}}
Let us now explore the Dirac's equation (where $\vec{\alpha}$ and $\beta$ are $4\times 4$ matrices)
\begin{equation}
i\hbar\frac{\partial\psi}{\partial t}=-i\hbar c\vec{\alpha}\cdot\vec{\nabla}\psi+\beta\,mc^2\psi\,,
\end{equation}
by writing the wavefunction $\psi$ as $\psi=R\,e^{iS/\hbar}$\,, where $S$ and $R$ can be seen as 4-components real functions. Equating the real and imaginary parts in the two sides of the resulting equation yield, respectively,
\begin{equation}
\frac{\partial S}{\partial t}+\left( c\vec{\alpha}\cdot\vec{\nabla}S+\beta\,mc^2\right)=0\,,
\qquad\qquad\frac{\partial R}{\partial t}+c\,\vec{\alpha}\cdot\vec{\nabla}R=0\,.
\end{equation}
Since the velocity in the Dirac formulation is given by $\vec{v}=c\vec{\alpha}$\,, then eq.(37)  will be
\begin{equation}
\left(\frac{\partial S}{\partial t}+\vec{v}\cdot\vec{\nabla}S\right)+\beta\,mc^2=0\,,
\qquad\qquad
\frac{\partial R}{\partial t}+\vec{v}\cdot\vec{\nabla}R=0\,.
\end{equation}
It is interesting to see that the quantum equation in Eq.(36) is reduced to two classical equations, Eq.(37) and (38). Equation (38) can be written as
\begin{equation}
\frac{d S}{d t}+\beta\,mc^2=0\,, \qquad\qquad \frac{d R}{d t}=0\,,
\end{equation}
or by using the relation  $\vec{v}=\frac{\vec{\nabla}S}{m}$ in eq.(38) yields
\begin{equation}
\frac{\partial S}{\partial t}+\left(\frac{(\nabla S)^2}{m}+\beta\,mc^2\right)=0\,,
\qquad\qquad
\frac{\partial R}{\partial t}+\frac{\vec{\nabla}S\cdot\vec{\nabla}R}{m}=0\,.
\end{equation}
The second equation in eq.(39) implies that the probability (amplitude) is time independent. The first equation in eq.(40) suggests that the Dirac total energy in the classical world is given by $E_D=\frac{p^2}{m}\pm  mc^2$. The first equation in Eq.(39) implies that $S\propto -\beta\,  mc^2t$\,, where the eigenvalues of $\beta$ are $\pm\,1$; and the second equation implies that $R=const.$

Equation (40) can be seen as representing two Hamilton-Jacobi equations with actions $S$ and ${\cal R}=\hbar R$, \emph{viz}.,
\begin{equation}
\frac{\partial S}{\partial t}+H_S=0\,,\qquad\qquad H_S=\frac{(\nabla S)^2}{m}+\beta\,mc^2\,,
\end{equation}
and
\begin{equation}
\frac{\partial {\cal R}}{\partial t}+H_R=0\,,\qquad\qquad H_R=\frac{\hbar}{m}\vec{\nabla}S\cdot\vec{\nabla}R\,.
\end{equation}
The two Hamiltonians in eq.(41) and (42) can be expressed as
\begin{equation}
H_S=\frac{p_S^2}{m}+\beta\,mc^2\,,\qquad\qquad H_R=\frac{\vec{p}_S\cdot\vec{p}_R}{m}\,,
\end{equation}
so that Hamilton's equations yield
\begin{equation}
\frac{dR}{dt}=\frac{p_S}{\lambda\, m}\,,\qquad\qquad \frac{dS}{dt}=\frac{2p_S p_\lambda}{m}\,,
\end{equation}
where $p_S=\rm const.$,  since $H_S$ is independent of  $S$, and $H_R$ is independent of $R$, respectively.

\section{\textcolor[rgb]{0.00,0.07,1.00}{Bohmian Klein-Gordon equation}}
 Let us now consider the  Klein-Gordon equation
\begin{equation}
 \frac{1}{c^2}\frac{\partial^2\phi}{\partial t^2}-\nabla^2\phi+\left(\frac{mc}{\hbar}\right)^2\phi=0\,,
 \end{equation}
and substitute instead $\phi=R\,e^{iS/\hbar}$. Separating the real and imaginary parts in the resulting equation, yields
\begin{equation}
\frac{1}{c^2}\,\ddot R-R''+\left(\frac{m^2c^2}{\hbar^2}+\frac{S'^2}{\hbar^2}-\frac{\dot S^2}{c^2\hbar^2}\right)R=0\,,
\end{equation}
and
\begin{equation}
\frac{1}{c^2}\,\ddot S-S''+2\left(\frac{1}{c^2}\frac{\dot R}{R}\frac{\dot S}{S}-\frac{R'}{R}\frac{S'}{S}\right)S=0\,.
\end{equation}
If we allow $R$ to satisfy the Klein-Gordon equation, then $cS'=\dot S$. This latter equation implies that $S$ satisfies a wave traveling at the speed of light. Conversely, if we allow $R$ to satisfy the wave equation traveling at the speed of light, then eq.(46) would imply that $S$ satisfies the equation, $\dot S^2=c^2S'\,^2+m^2c^4$, which upon using the Hamilton-Jacobi equation, $E=-\frac{\partial S}{\partial t}=-\dot S$ and $\vec{p}=\vec{\nabla}S=S'$ yields $E^2=c^2p^2+m^2c^4$ that is the Einstein relativistic energy-momentum equation. A massless particle satisfies the two wave equations $c R'=\dot R$ and $cS'=\dot S$.  Interestingly, the amplitude and the phase of the wavefunction of a massless particle satisfy the wave equation traveling at the speed of light.

Equations (46) and (47) suggest that the particle consists of two subparticles whose fields satisfy the Klein-Gordon equation with masses
\begin{equation}
M^2_{R}=m^2+\frac{S'^2}{c^2}-\frac{\dot S^2}{c^4}\,,\qquad\qquad
M^2_{S}=\frac{2\hbar^2}{c^2}\left(\,\frac{1}{c^2}\frac{\dot R\dot S}{RS}-\,\frac{R'S'}{RS}\right)\,.
\end{equation}
The mass of the $R$ field depends solely on $S$ fields, but that due to $S$ field depends on $S$ and $R$ fields and their spatial and temporal derivatives. Hence, the mass of each subfield depends on the interaction between these subfields.
Thus, the field $\phi$ is composed of these two subfields (eigen functions). This is manifested in writing $\phi$ as $\phi=U+iW$, where $U=R\cos(S/\hbar)$ and $W=R\sin(S/\hbar)$, where $U$ and $W$ satisfy the Klein-Gordon equation.

\section{\textcolor[rgb]{0.00,0.07,1.00}{Bohmian unified quantum  wave equation}}
 Let us now apply the wavefunction representation,  $\psi_0=R\exp\,(iS/\hbar)\,,$ in the unified quantum wave equation \textcolor[rgb]{0.00,0.07,1.00}{\cite{uqm}}
 \begin{equation}
 \frac{1}{c^2}\frac{\partial^2\psi_0}{\partial t^2}-\nabla^2\psi_0+\frac{2m}{\hbar}\frac{\partial\psi_0}{\partial t}+\left(\frac{mc}{\hbar}\right)^2\psi_0=0\,.
 \end{equation}
Equating the real and imaginary parts in the two sides of the resulting equation to each other, yields
\begin{equation}
\frac{\ddot R}{c^2}-R\,''+\frac{2m}{\hbar}\,\dot R+\left(\frac{m^2c^2}{\hbar^2}-\frac{\dot S^2}{c^2\hbar^2}+\frac{S'^2}{\hbar^2}\right)\,R=0\,
\end{equation}
and
\begin{equation}
\frac{\ddot S}{c^2}-S''+\frac{2m}{\hbar}\,\dot S+2\left(\frac{1}{c^2}\frac{\dot R}{R}\frac{\dot S}{S}-\frac{\,\,\,R\,'}{R}\frac{S'}{S}\,\right)S =0\,.
\end{equation}
It is apparent that if we assume  $R$  satisfy eq.(49), then one finds $\dot S=cS'\,. $
This implies that $S$ satisfies a wave equation with zero mass, \emph{viz}., $\frac{\ddot S}{c^2}-S''=0$, while $R$ satisfies a dissipative wave equation with non-zero mass. 
Thus, the phase information travels at the speed of light. Therefore, the wavefunction $\psi_0$ contains two kinds of waves. Hence, particles and waves are hybrid (coexist). The two waves are coupled, as evident from eqs.(50) and (51).  However, if we assume that $R$ satisfy a modulated Klein-Gordon equation (telegraph equation), \emph{i.e.},
\begin{equation}\frac{\ddot R}{c^2}-R\,''+\frac{2m}{\hbar}\,\dot R=0\,
\end{equation}
then eq.(50) yields
\begin{equation}
\frac{\dot S^2}{m^2c^4}-\frac{S'^2}{m^2c^2}=1 \,,
\end{equation}
which is an equation of a hyperbola with coordinates, $\dot S$ and $S\,'$. Thus, the phase evolves dynamically and, at any given time,  is controlled by the hyperbola equation.

Equations (50) and (51) indicate that in the new functions, $S$ and $R$,  the quantum particle appears to be consisting of two interacting subparticles satisfying eq.(49). The masses of these subparticles are, respectively,
\begin{equation}
M^2_R=m^2-\frac{\dot S^2}{c^4}+\frac{S'^2}{c^2}\,,\qquad\qquad M^2_S=  \frac{2\hbar^2}{c^4}\frac{\dot R}{R}\frac{\dot S}{S}-\frac{2\hbar^2}{c^2}\frac{\,\,\,R\,'}{R}\frac{S'}{S}
\end{equation}
It is important to compare this formulation with Schr\"{o}dinger one. In Schr\"{o}dinger formulation, the wave amplitude  $R$ is not coupled with the phase, $S$. Moreover,  $S$  satisfies the quantum Hamilton-Jacobi equation but $R$ doesn't satisfy Schr\"{o}dinger equation. In the present formulation while $R$ is a decaying oscillatory function of time satisfying the telegraph equation, the phase satisfies a wave equation traveling at the speed  of light. Hence, the phase information is dispatched everywhere at the speed of light. Moreover, the field $\psi_0$ is composed of two subfields, $U$ and $W$, such that $\psi_0=U+iW$, where $U=R\cos(S/\hbar)$ and $W=R\sin(S/\hbar)$. Note that $U$ and $W$ satisfy the wave equation in eq.(49)

Let us now consider the special (stationary) case, when $\dot R=0$ and $\dot S=0$, and substitute them in eqs.(50) and (51) to obtain
\begin{equation}
\hbar^2c^2\frac{\nabla^2R}{R}=m^2c^4+c^2S'^2\,,\qquad\qquad S'=AR^{-2}\,,\qquad\qquad A=\rm const. \,.
\end{equation}
Hence, the quantum potential defined in Schrodinger case can be expressed by
\begin{equation}
V_Q=-\frac{\hbar^2}{2m}\frac{\nabla^2R}{R} =-\frac{mc^2}{2}-\frac{A^2}{2mR^4}\,,
\end{equation}
which is an attractive potential.
Since as $r\rightarrow\infty$\,, the probability, $R^2\rightarrow 0$, the quantum potential $V_Q\rightarrow-\infty$. This indicates that the influence of the quantum potential (force) is present even at large scale. The quantum potential may be responsible for the splitting (creation/annihilation) of the $\psi_0$ field into $U$ and $W$ fields each having a mass of $mc^2$. These two fields oscillate internally so as to endow the particle its quantum effects (spin, jittery motion, etc.).

\section{\textcolor[rgb]{0.00,0.07,1.00}{Concluding remarks}}
We have  revisited  in this paper the Bohmian quantum mechanics by considering that the continuity equation can be set to represent an additional Hamilton-Jacobi equation; and solve the two Hamilton equations  treating $S$ and $R$ as two generalized co-ordinates. In doing so, a new potential, called the phase potential is obtained, that is proportional to the curvature of $S$. This is a classical potential that doesn't incorporate the Planck's constant. The phase  potential gives rise to a viscosity force whose coefficient of viscosity is directly proportional to the phase. Furthermore, we apply the Bohmian recipe  of quantum mechanics to the unified quantum wave equation that  we have recently derived.  Besides, we have shown that if $R$ satisfies the telegraphic de Broglie equation (undistorted telegraph equation), then $S$ satisfies the wave equation traveling at the speed of light. However, if we assume $R$ satisfy the telegraph equation, then  $S$ satisfies the Einstein relativistic energy-momentum equation expressed in terms of the momentum $\vec{p}=\vec{\nabla}S$ and energy $E=-\frac{\partial S}{\partial t}$. Applying the Bohmian recipe to the Dirac's equation shows that $R^2$ is conserved and the rate of change of the phase density, $\frac{d}{dt}(R^2S)=-\beta mc^2R^2$. It is worth to mention that our Bohmian treatment for Dirac equation yields two classical equations, whereas Schrodinger, Klein-Gordon and the unified quantum wave equations do split into additional quantum equations.

Furthermore, it is shown that the functions $R$ and $S$ represent two subfields from which the quantum particle is formed, and the interaction between them provides their masses.

\section{\textcolor[rgb]{0.00,0.07,1.00}{Acknowledgements}}
I would like to thank the referee for his/her critical comments and suggestions.


\end{document}